\documentclass[conference,a4paper]{IEEEtran}
\IEEEoverridecommandlockouts

\usepackage{xcolor}
\usepackage{balance}

\ifCLASSINFOpdf
  \usepackage[pdftex]{graphicx}
\else
  \usepackage[dvips]{graphicx}
\fi
\ifCLASSOPTIONcompsoc
 \usepackage[caption=false,font=normalsize,labelfont=sf,textfont=sf]{subfig}
\else
 \usepackage[caption=false,font=footnotesize]{subfig}
\fi

\usepackage{graphicx} 
\graphicspath{{Fig/}}

\usepackage{graphicx}
\usepackage{amsmath,bbm,amssymb,amsfonts,amstext,amsopn,sansmath}
\usepackage{cite}
\usepackage{balance}
\usepackage{url}
\usepackage{epsfig}
\usepackage{setspace}
\usepackage{stmaryrd}
\usepackage{psfrag}	
\usepackage{multirow}
\usepackage{makecell}
\usepackage{float}
\usepackage[process=auto]{pstool}
\usepackage{etoolbox}
\usepackage{algorithm}
\usepackage{algorithmic}
\usepackage{flushend}
\allowdisplaybreaks
\usepackage{textcomp}

\newcommand{\e}{\mathsf{e}}
\newcommand{\jj}{\mathsf{j}}
\newcommand{\Herm}{\mathsf{H}}
\newcommand{\Trans}{\mathsf{T}}
\newcommand{\x}{\mathsf{x}}
\newcommand{\y}{\mathsf{y}}
\newcommand{\z}{\mathsf{z}}

\newcommand{\bx}{\mathbf{x}}
\newcommand{\by}{\mathbf{y}}
\newcommand{\bn}{\mathbf{n}}
\newcommand{\bH}{\mathbf{H}}
\newcommand{\ba}{\mathbf{a}}
\newcommand{\bu}{\mathbf{u}}
\newcommand{\bd}{\mathbf{d}}
\newcommand{\bR}{\mathbf{R}}
\newcommand{\bM}{\mathbf{M}}
\newcommand{\bU}{\mathbf{U}}
\newcommand{\bV}{\mathbf{V}}

\newcommand{\bI}{\mathbf{I}}
\newcommand{\bX}{\mathbf{X}}
\newcommand{\bh}{\mathbf{h}}
\newcommand{\bbf}{\mathbf{f}}

\newcommand{\bGamma}{\boldsymbol{\Gamma}}
\def\bomega{\boldsymbol{\omega}}
\def\bPsi{\boldsymbol{\Psi}}
\def\bzero{\boldsymbol{0}}

\newcommand{\tx}{\mathrm{tx}}

\def\Cset{\mathbb{C}}
\def\Rset{\mathbb{R}}
\newcommand{\Ex}{\mathbb{E}}
\newcommand{\Var}{\mathbb{V}}

\newcommand{\tr}{\mathrm{tr}}
\newcommand{\dd}{\mathrm{d}}
\newcommand{\sgval}{\mathsf{sg\underline\_val}}

\def\LOS{\mathrm{LOS}}
\def\nLOS{\mathrm{nLOS}}
\def\tx{\mathrm{tx}}
\def\rx{\mathrm{rx}}
\def\vect{\mathrm{vec}}
\def\sinc{\mathrm{sinc}}
\def\iid{\mathrm{iid}}
\def\PL{\mathrm{PL}}
\def\SF{\mathrm{SF}}
\def\CP{\mathrm{P}}
\def\BL{\mathrm{BL}}
\def\UE{\mathrm{UE}}
\def\thr{\mathrm{thr}}
\def\tile{\mathrm{tile}}

\def\sCN{\mathcal{CN}}
\def\sMCN{\mathcal{MCN}}
\def\sM{\mathcal{M}}
\def\sV{\mathcal{V}}

\newtheorem{lem}{Lemma}
\newtheorem{remk}{Remark}

\newtoggle{OneColumn}
\toggletrue{OneColumn}

\newtoggle{arXiv}
\togglefalse{arXiv}

\begin{document}
	 
%
\title{\huge Impact of Channel Models on Performance Characterization of RIS-Assisted Wireless Systems \vspace{-0.2cm}}

\author{Vahid Jamali$^{\ast,\diamond}$, Walid Ghanem$^{\star}$, Robert Schober$^{\star}$, and H. Vincent Poor$^{\diamond}$\\
	$^{\ast}$Technical University of Darmstadt, 
	$^{\diamond}$Princeton University,
	$^{\star}$Friedrich-Alexander-Universit\"at Erlangen-N\"urnberg \vspace{-0.3cm}
\thanks{This work was supported in part by the LOEWE initiative (Hesse, Germany) within
	the emergenCITY center, the Federal Ministry of Education and Research (BMBF, Germany) through the
	6G Research and Innovation Cluster 6G-RIC, and the U.S. National Science Foundation under Grant CNS-2128448.}}




       \maketitle

\begin{abstract}
	The performance characterization of communication systems assisted by large reconfigurable intelligent surfaces (RISs) significantly depends on the adopted models for the underlying channels. Under unrealistic channel models,  the system performance may be over- or under-estimated which yields inaccurate conclusions for the system design. In this paper, we review five channel models that are chosen to progressively improve the modeling accuracy for large RISs. For each channel model, we highlight the underlying assumptions, its advantages, and its limitations. We  compare the system performance under the aforementioned channel models using  RIS configuration algorithms from the literature and a new scalable algorithm proposed in this paper specifically for the configuration of extremely large RISs.
\end{abstract}


\section{Introduction}

Reconfigurable intelligent  surfaces (RISs) have been extensively studied in the literature over the past few years as a means to realize the emerging concept of a programmable radio environment \cite{di2019smart,yu2021smart}.  Due to their cost- and energy-efficiency, \textit{extremely large} passive RISs can be manufactured; see e.g., \cite{tang2020wireless} for initial proof-of-concept implementations containing up to several thousands RIS elements. In fact, it has been shown in the literature that for most typical scenarios, passive RISs \textit{must} be quite large  to be able to establish a sufficiently strong link~\cite{najafi2020intelligent}.

Most works in the literature have evaluated the performance of RIS-assisted communication systems assuming that the involved channel coefficients are independent and identically distributed (i.i.d.) and follow Rayleigh or Rician fading models \cite{wu2019intelligent,han2019large,yu2020robust}. The underlying assumptions for these models include a sufficiently-large element spacing, a small-to-moderate antenna aperture size, and an infinitely rich (and isotropic) scattering environment, which all may be violated for typical use cases of  RIS. In particular, in \cite{bjornson2020rayleigh}, it has been shown that for a rich scattering environment, the channel coefficients are  correlated even for the well-known case of half-wavelength element spacing. Moreover, the richness of the scattering channel is a relative quantity that depends on the antenna aperture size. In particular, since in a given environment, the number of channel paths is fixed regardless of the antenna aperture size, a channel that comprises a reasonably  large number of propagation paths may be considered rich for a small aperture but eventually, as the aperture size grows, the assumption of rich scattering breaks down \cite{najafi2020intelligent}. The channel matrices for large RISs feature certain structures (including a low-rank property) that cannot be captured only via correlation. Furthermore, for  extremely large RISs, the transmitter (Tx) and receiver (Rx) may fall in the near-field of the RIS where the wavefront curvature cannot be neglected across the entire RIS \cite{alexandropoulos2022near,ramezani2022near}.

The importance of adopting a reasonable channel model for performance evaluation of RIS-assisted communication systems is two-folded. On the one hand, for an unrealistic channel model,  the system performance may be over- or under-estimated which yields inaccurate conclusions for the system design, e.g., a significant performance improvement may seem achievable by a small RIS when an unrealistic channel model is assumed. On the other hand, the algorithms developed for small RISs may not be straightforwardly applicable to large RISs due to their limited scalability, channel state information (CSI) requirements, etc. Therefore,  fundamental questions regarding channel models for RIS-assisted wireless systems include: What is the impact of different channel models on the system performance? and which models are applicable for a given set of system parameters? To the best of the authors' knowledge, a performance comparison of RIS-assisted communication systems from the perspective of the underlying channel models has not been reported in the literature, yet, which constitutes the main objective of the present study. 

In this paper, we review five channel models that are chosen to progressively improve the modeling accuracy for large RISs. For each channel model, we highlight the underlying assumptions, its advantages, and its limitations. We first compare the system performance under the aforementioned channel models for two algorithms proposed in \cite{wu2019intelligent}. Since these algorithms rely on semidefinite programming (SDP), their computational complexity scales cubically in the number of RIS elements, denoted by $Q$, which becomes prohibitive for large RISs. Therefore, we propose a scalable algorithm with linear complexity in $Q$ which enables us to compare the system performance under different channel models for extremely large RISs, too. Our simulation results quantify the performance over-/under-characterization by each model with respect to (w.r.t.) other considered models and reveal interesting insights for the system design. For instance, while for most scenarios, an isotropic scattering environment yields a higher system performance, the proposed RIS configuration algorithm is able to achieve a higher performance in the finite scattering case for very large RISs by explicitly exploiting the underlying geometric characteristics of wave propagation.

\textit{Notation:} Bold capital and small letters are used to denote matrices and vectors, respectively.  $(\cdot)^\Trans$, $(\cdot)^\Herm$, $\vect(\cdot)$, and $\otimes$ denote the  transpose, Hermitian, and vectorization operators, and the Kronecker product, respectively. Moreover, $\boldsymbol{0}_n$ and $\boldsymbol{0}_{n\times m}$ denote a column vector of size $n$ and a matrix of size $n\times m$, respectively, whose elements are all zeros, and $\mathbf{I}_n$ is the $n\times n$ identity matrix.  $[\bX]_{m,n}$ and $[\bx]_{n}$ denote the element in the $m$th row and $n$th column of matrix $\bX$ and the $n$th entry of vector $\bx$, respectively.  $\Ex\{\cdot\}$ and $\Var\{\cdot\}$ represent expectation and variance, respectively. $\mathcal{CN}(\boldsymbol{\mu},\boldsymbol{\Sigma})$ denotes a complex Gaussian random vector with mean vector $\boldsymbol{\mu}$ and covariance matrix $\boldsymbol{\Sigma}$. Moreover,  $\bX$ is said to follow a complex Gaussian matrix distribution $\mathcal{MCN}(\bM,\bU,\bV)$, if $\vect(\bX)\sim\mathcal{CN}(\vect(\bM),\bU\bV)$.
Finally, $\mathbb{R}$ and $\mathbb{C}$ represent the sets of real and complex numbers, respectively. 

\section{System Model}

We consider a narrow-band downlink communication between a base station (BS) and $N_{\UE}$ user equipments (UEs) assisted by an RIS. The signal received at the $k$th UE is~given~by
\begin{IEEEeqnarray}{ll}\label{Eq:IRSbasic}
	\by_k = \big(\bH_{d,k} + \bH_{r,k} \bGamma \bH_t \big) \bx +\bn_k,\quad k=1,\dots,N_{\UE},
\end{IEEEeqnarray}
where $\bx\in\Cset^{N_t}$ is the transmit signal vector with $N_t$ denoting the number of BS antennas, $\by_k\in\Cset^{N_r}$ is the received signal vector at the $k$th UE with $N_r$ being the number of UE antennas, and  $\bn_k\in\Cset^{N_r}$ represents the additive while Gaussian noise (AWGN) at the $k$th UE, i.e., $\bn_k\sim\sCN(\bzero_{N_r},\sigma_n^2\bI_{N_r})$, where $\sigma_n^2$ is the noise power. Moreover,  $\bH_{d,k}\in\Cset^{N_r\times N_t}, \bH_t\in\Cset^{Q\times N_t}$, and $\bH_{r,k}\in\Cset^{N_r\times Q}$ denote the BS-UE, BS-RIS, and RIS-UE channel matrices, respectively, where $Q$ is the number of RIS unit cells. Furthermore, $\bGamma\in\Cset^{Q\times Q}$ is a diagonal matrix with main diagonal entries $\Gamma_q=\Omega_q\e^{\jj\omega_q}$ denoting the reflection coefficient applied by the $q$th RIS unit cell comprising phase shift $\omega_q$  and reflection amplitude $\Omega_q$. The design and performance of the RIS-assisted communication systems significantly depend on the channel models assumed for channel matrices $\bH_{d,k}$,  $\bH_t$,  and $\bH_{r,k}$, which will be discussed in detail next.

\section{Channel Models for IRS-assisted Systems}

In this section, we review the most widely-adopted  channel models in the literature. We drop subscripts $d$, $t$,  $r$, and $k$ for notational simplicity, explain the channel models for a general matrix $\bH\in\Cset^{N_\rx\times N_\tx}$ corresponding to $N_\tx$ transmit antenna elements and $N_\rx$ receive antenna elements, and wherever necessary, explicitly refer to the  BS-UE, BS-RIS, and RIS-UE channels.

\subsection{I.I.D. Rayleigh Fading}\label{sec:Rayleigh}

The  Rayleigh fading model   assumes that the entries of channel matrices are i.i.d. complex Gaussian random variables (RVs),~i.e., 
\begin{IEEEeqnarray}{ll}\label{Eq:Rayleigh}
	[\bH]_{m,n} \sim \sCN\big(0,\bar{h}_{\CP}\big)
\end{IEEEeqnarray}
and $[\bH]_{m,n}$ and $[\bH]_{m',n'}$ are independent for any $(m,n)\neq (m',n')$. Here, $\bar{h}_{\CP}$  is the channel power and captures the impact of the distance-dependent channel pathloss, denoted by $\bar{h}_{\PL}$, the loss due to blockage, denoted by $\bar{h}_{\BL}$, and the large-scale shadow fading, denoted by $\bar{h}_{\SF}$. For i.i.d. Rayleigh fading, there exists no structure among the entries of the channel matrix. The underlying assumptions for this model are that the antenna elements are well separated (i.e., more than half a wavelength)  there exist \textit{infinitely} many multipath components in an \textit{isotropic} scattering environment~\cite{bjornson2020rayleigh}.

\subsection{I.I.D. Rician Fading}\label{sec:Rician}

It is usually desirable to deploy the RIS such that there exist line-of-sight (LOS) links for the BS-RIS and RIS-UE channels. In this case, Rician fading is a more accurate small-scale fading model to describe BS-RIS and RIS-UE channels \cite{wu2019intelligent,han2019large}, i.e.,
\begin{IEEEeqnarray}{ll}\label{Eq:Rician}
	\bH = \sqrt{\frac{K}{1+K}}\bH^\LOS + \sqrt{\frac{1}{1+K}}\bH^{\nLOS},
\end{IEEEeqnarray}
where $\bH^\LOS$ and $\bH^\nLOS$ are the LOS and non-LOS components of $\bH$, respectively, and $K$ denotes the K-factor and determines the relative power of the LOS component compared to the  non-LOS components of the channel. The non-LOS channel matrix $\bH^\nLOS$ is modeled similar to \eqref{Eq:Rayleigh}. The LOS channel matrix $\bH^\LOS$ depends on the antenna array deployments and the relative angles of the Tx and Rx arrays. Assuming the far-field regime, $\bH^\LOS$ is modeled as
\begin{IEEEeqnarray}{ll}\label{Eq:LOS}
	\bH^\LOS = c\, \ba_{\rx}(\bPsi_\rx)\ba_{\tx}^\Herm(\bPsi_\tx),
\end{IEEEeqnarray}
where $c=\sqrt{\bar{h}_{\CP}}$ is the channel attenuation factor of the LOS link, and $\ba_{\tx}(\cdot)\in\Cset^{N_\tx}$ and $\ba_{\rx}(\cdot)\in\Cset^{N_\rx}$ denote the Tx and Rx array steering vectors, respectively. Moreover, $\bPsi_\tx=(\theta_\tx,\phi_\tx)$ is the Tx angle-of-departure (AoD), i.e., the direction of the Rx defined in the Tx array coordinate system, where $\theta_\tx$ and $\phi_\tx$ denotes the elevation and azimuth angles, respectively. Similarly, $\bPsi_\rx=(\theta_\rx,\phi_\rx)$ is Rx angle-of-arrival (AoA), i.e.,  the direction of the Tx  defined in the Rx array coordinate system. The array steering vector depends on the array manifold, i.e., the positions of the antenna elements, denoted by $\bu_n\in\Rset^3$, $n=1,\dots,N$, with $N$ being the number of array elements, and is given~by
\begin{IEEEeqnarray}{ll}\label{Eq:steering}
	\ba(\bPsi) = \left[\e^{\jj\kappa\bd^\Trans\!(\bPsi)\bu_1},\dots,\e^{\jj\kappa\bd^\Trans\!(\bPsi)\bu_N}\right]^\Trans,
\end{IEEEeqnarray}
where $\kappa=\frac{2\pi}{\lambda}$  is the wave number, $\lambda$ denotes the wavelength, and $\bd(\bPsi)\in\Rset^3$ is a unit vector pointing towards the direction of $\bPsi$. Depending on the specific adopted array manifold and the choice of coordinates system, \eqref{Eq:steering} can be often further simplified. For instance, assuming a uniform planar array (UPA) located in the $\y-\z$ plane and consisting of $N_\y$ and $N_\z$ antenna elements (i.e., $N_\y N_\z=N$) spaced by $d_\y$ and $d_\z$ along the $\y$- and $\z$-axes, indexed by $n_\y=0,\dots,N_\y-1$ and $n_\z=0,\dots,N_\z-1$, respectively, we obtain $\bu_{(n_\y,n_z)}=[0,n_\y d_\y, n_\z d_\z]^\Trans$  and $\bd(\bPsi)=[\cos(\theta) \cos(\phi), \cos(\theta) \sin(\phi), \sin(\theta)]^\Trans$, where the first antenna element $(n_\y,n_\z)=(0,0)$ is assumed to be located at the origin. Therefore, $[\ba(\bPsi)]_n$ corresponding to antenna element $n$ (parameterized by $(n_\y,n_z)$ in the UPA) is given by
\begin{IEEEeqnarray}{ll}\label{Eq:UPA}
	[\ba(\bPsi)]_n = \e^{\jj\kappa \left[d_\y \cos(\theta) \sin(\phi) n_\y + d_\z \sin(\theta) n_\z \right]}.
\end{IEEEeqnarray}

\begin{remk}
	The steering vector of the UPA, $\ba(\bPsi)\in\Cset^{N}$, can be written as the Kronecker product of the steering vectors of uniform linear arrays (ULAs) along the $\y$- and $\z$-axes, denoted by $\ba_\y(\bPsi)\in\Cset^{N_\y}$ and $\ba_\z(\bPsi)\in\Cset^{N_\z}$, respectively, as follows
	\begin{IEEEeqnarray}{ll}\label{Eq:UPAkron}
		\ba(\bPsi) = \ba_\y(\bPsi) \otimes \ba_\z(\bPsi),
	\end{IEEEeqnarray}
	where 
	\begin{IEEEeqnarray}{ll}\label{Eq:ULA}
		[\ba_\y(\bPsi)]_{n_\y} \!=\! \e^{\jj\kappa d_\y \cos(\theta) \sin(\phi) n_\y}
		\,\,\text{and}\,\,
		[\ba_\z(\bPsi)]_{n_\z} \!=\! \e^{\jj\kappa d_\z \sin(\theta) n_\z}.\quad\,
	\end{IEEEeqnarray}
\end{remk}

\subsection{Correlated Rayleigh Fading}

The Rayleigh fading model introduced in Section~\ref{sec:Rayleigh} stems from the fact that for rich scattering environments, the effective channel gain originating from an infinitely large number of multipath components follows a Gaussian distribution by the central limit theorem. However, for a given array structure, the entries of the channel matrix may not necessarily be i.i.d.  RVs. In the context of RIS-assisted communication, a corresponding correlation model was derived in  \cite{bjornson2020rayleigh} for a single-input multiple-output (SIMO) channel. In the following, we generalize this result to multiple-input multiple-output (MIMO) channels. In particular, the  non-LOS channel is assumed to comprise $L\to\infty$ channel paths, i.e.,
\begin{IEEEeqnarray}{ll}\label{Eq:nLOS}
	\bH^\nLOS = \frac{1}{\sqrt{L}} \sum_{\ell=1}^L c^{(\ell)}\ba_{\rx}(\bPsi_\rx^{(\ell)})\ba_{\tx}^\Herm(\bPsi_\tx^{(\ell)}),
\end{IEEEeqnarray}
where superscript $\ell$ denotes the $\ell$th channel path. We adopt the same assumptions as made in  \cite{bjornson2020rayleigh}, namely
\begin{itemize}
	\item[A1)] $\frac{c^{(\ell)}}{\sqrt{L}}$ denotes the complex signal attenuation of the $\ell$th channel path. Moreover, $c^{(\ell)}$, $\forall \ell$, are i.i.d. RVs with  mean zero and variance $\sigma_c^2=\bar{h}_{\CP}$.
	\item[A2)] The Tx and Rx arrays are deployed in an isotropic rich scattering environment where the multipath components are uniformly distributed over the half-space in front of them. Therefore, angles $\bPsi_\rx^{(\ell)},\bPsi_\tx^{(\ell)}$  are i.i.d. RVs characterized by the following probability density function (PDF) 
	\begin{IEEEeqnarray}{ll}\label{Eq:Angles}
		f(\bPsi) = \frac{\cos(\theta)}{2\pi},\quad 
		\theta\in\left[-\frac{\pi}{2},\frac{\pi}{2}\right], \phi\in\left[-\frac{\pi}{2},\frac{\pi}{2}\right]. 
	\end{IEEEeqnarray}
\end{itemize}

\begin{lem}\label{Lem:Correlation}
	Under assumptions A1 and A2, and assuming $L\to\infty$, the statistical distribution of $\bH^\nLOS$ in  \eqref{Eq:nLOS} approaches the following correlated matrix Gaussian distribution:
	\begin{IEEEeqnarray}{ll}\label{Eq:matGaussian}
		\bH^\nLOS \sim \sMCN(\bzero_{N_\rx,N_\tx},\sigma_c\bR_\rx,\sigma_c\bR_\tx)
	\end{IEEEeqnarray}
	where $\bR_\tx\in\Cset^{N_\tx\times N_\tx}$  and
	$\bR_\rx\in\Cset^{N_\rx \times N_\rx}$ denote the transmit and receive correlation matrices, respectively, which are given by
	\begin{IEEEeqnarray}{ll}\label{Eq:CorrelationMatrix}
	[\bR_s]_{m,n} = \sinc\big(\kappa\|\bu_{s,m}-\bu_{s,n}\|\big),
	\quad s\in\{\tx,\rx\},
\end{IEEEeqnarray}
where $\sinc(x)=\frac{\sin(x)}{x}$ is the sinc function and $\bu_{s,n}$ denotes the position of the $n$th antenna at node $s\in\{\tx,\rx\}$.
\end{lem}
\begin{IEEEproof}
	The proof follows the same steps as the corresponding proof  in \cite{bjornson2020rayleigh} but is included in Appendix~A for completeness. 
\end{IEEEproof}

As can be seen from \eqref{Eq:CorrelationMatrix}, the correlation among two antenna elements is only zero if their distance is a multiple integer of $\lambda/2$. Therefore, for a UPA RIS with $\lambda/2$ element spacing,  the correlation cannot be zero for all element pairs (zero correlation is guaranteed for elements on the same row or column, though).

\begin{remk}
	The matrix Gaussian distribution $\sMCN(\bzero,\sigma_c\bR_\rx,\sigma_c\bR_\tx)$ can be generated as follows~\cite[Ch.~2]{gupta2018matrix}
	\begin{IEEEeqnarray}{ll}\label{Eq:matGaussianGenerate}
		\bH^\nLOS = \bar{\bR}_\rx\bH^\iid \bar{\bR}_\tx
	\end{IEEEeqnarray}
	where $\bar{\bR}_s\in\Cset^{N_s\times N_s},\,\,s\in\{\tx,\rx\}$ is found from the decomposition $\bR_s = \bar{\bR}_s\bar{\bR}_s^\Trans$ and the entries of $\bH^\iid\in\Cset^{N_\rx\times N_\tx}$ are i.i.d. complex Gaussian RVs distributed as $\sCN(0,\sigma_c^2)$. Alternatively,  matrix $	\bH^\nLOS \sim \sMCN(\bzero,\sigma_c\bR_\rx,\sigma_c\bR_\tx)$ can be generated as follows
	\begin{IEEEeqnarray}{ll}\label{Eq:vecGaussian}
		\vect\left(\bH^\nLOS\right) \sim \sCN(\bzero_{N_\rx N_\tx},\sigma_c^2\bR_\rx\otimes\bR_\tx).
	\end{IEEEeqnarray}
\end{remk}

\subsection{Low-rank Geometric Channel Model}

The assumption of rich scattering made so far is valid when the number of channel paths is significantly large compared to the number of antenna elements. However, this condition eventually breaks down as the aperture size grows, particularly at high carrier frequencies. This is because as the carrier frequency increases, on the one hand, the number of  antenna elements has to increase to cope with the larger path loss, and on the other hand, the number of effective channel scatters decreases due to the higher scattering losses and atmospheric absorption at higher frequencies. Nevertheless, the impact of finiteness of the number of channel paths can be noticeable even at lower frequencies when the antenna aperture is extremely large (as in the case of large RISs).    Thereby, a geometric path-based channel model is often adopted \cite{najafi2020intelligent,jaeckel2016quadriga} which is described in the following. In practice, the signal sent by the Tx will interact with objects in the environment (e.g.,  building walls, office furnishings, etc.) before reaching the Rx \cite{jaeckel2016quadriga}. Depending on the roughness of the object surface, it may generate not only a single reflection path but a collection of scattered paths that are  spatially confined. To model this behavior, the channel is characterized by a collection of channel scattering objects each generating a number of channel paths that are closely spaced in the angular domain. This leads to the following model \cite{jaeckel2016quadriga}
\begin{IEEEeqnarray}{ll}\label{Eq:nLOSlowrank}
	\bH^\nLOS \!=\! \frac{1}{\sqrt{LR}} \sum_{\ell=1}^L \sum_{r=1}^R g^{(\ell)}\e^{\jj\psi^{(\ell,r)}}\!\! \ba_{\rx}(\bPsi_\rx^{(\ell,r)})\ba_{\tx}^\Herm(\bPsi_\tx^{(\ell,r)}), \quad\,\,
\end{IEEEeqnarray}
where $R$ is the number of sub-paths in each cluster,  $g^{(\ell)}$ is the path attenuation coefficient for the $\ell$th scatter cluster and $\psi^{(\ell,r)}$ is the respective phase of the $r$th sub-path in cluster $\ell$. The values of $g^{(\ell)}$ and $\psi^{(\ell,r)}$ depend on the properties of the scattering object and the length of the propagation path, which for a given scenario are deterministic quantities. However, the way that the sub-paths add up can be either constructive or destructive which is physically the origin of  multipath fading. In fact, under the assumption that the  angular spread of paths inside each cluster is small, i.e.,  $\bPsi_s^{(\ell,r)}\approx\bPsi_s^{(\ell)},\,\,\forall\ell, s\in\{\tx,\rx\}$, the model in \eqref{Eq:nLOSlowrank} reduces to that in \eqref{Eq:nLOS} with fading coefficient $c^{(\ell)}=\frac{g^{(\ell)}}{\sqrt{R}}\sum_{r=1}^R \e^{\jj\psi^{(\ell,r)}}$ where $\Ex\{g^{(\ell)}\}=0$ and $\Var\{g^{(\ell)}\}=\bar{h}_{\CP}$. In other words, the cluster sub-paths are used in \eqref{Eq:nLOSlowrank} to emulate multipath fading.

\subsection{Geometric Channel Model for Large RISs}

Next, we present a general geometric channel model that is suitable for large RISs. In particular, we generalize \eqref{Eq:nLOSlowrank} to account for the following considerations:
\begin{itemize}
	\item \textbf{Wavefront curvature:} Due to the large aperture of practical passive RISs, the Tx, the Rx, and channel scatters may not be necessarily within its Fraunhofer distance. Therefore, the linear phase progression assumed in \eqref{Eq:LOS}, \eqref{Eq:nLOS}, and \eqref{Eq:nLOSlowrank} is not valid since the wavefront curvature (due to the spherical wave propagation) across the RIS cannot~be~neglected.
	\item \textbf{Amplitude taper:} Essentially, for large RISs, each channel path impinges on (reflects from) different points on the surface with  slightly \textit{different angles}. This leads to an amplitude variation across the RIS due to an angle-dependent unit-cell factor and unit-cell polarization.
\end{itemize}

For notational simplicity, we characterize the channel between transmit antenna element $n$ and receive antenna element $m$ due to the contribution of path $r$  in scattering cluster $\ell$ by two quantities, namely the propagation distance $d^{(\ell,r,m,n)}$ and transmit and receive channel AoD $\bPsi_{\tx}^{(\ell,r,m,n)}$ and AoA $\bPsi_{\rx}^{(\ell,r,m,n)}$, respectively. For the LOS component, we drop  indices $\ell$ and $r$. Based on these notations, the considered channel model reads \cite{jaeckel2016quadriga}
\begin{IEEEeqnarray}{ll}\label{Eq:proposed}
	[\bH^\nLOS]_{m,n} \!\!=\!\!  \frac{1}{\sqrt{LR}}\sum_{\ell=1}^L \sqrt{\bar{h}_{\CP}}   \nonumber\\
	\quad \times \sum_{r=1}^R\bbf^\Trans_{\rx}\big(\bPsi_{\rx}^{(\ell,r,m,n)}\big) \bM^{(\ell)} \bbf_{\tx}\big(\bPsi_{\tx}^{(\ell,r,m,n)}\big) 
	\e^{\jj \kappa d^{(\ell,r,m,n)}}\!\!\!\quad\quad \IEEEyesnumber\IEEEyessubnumber\\
	{[\bH^{\LOS}]_{m,n}} \!\!=\!\!   \sqrt{\bar{h}_{\CP}}  \bbf^\Trans_{\rx}\big(\bPsi_{\rx}^{(m,n)}\big) \bM \bbf_{\tx}\big(\bPsi_{\tx}^{(m,n)}\big) 
	\e^{\jj \kappa d^{(m,n)}}\!\!\!.\quad\,\,\,\,\,\,\IEEEyessubnumber
\end{IEEEeqnarray}
In \eqref{Eq:proposed}, $\bbf_{\tx}\big(\bPsi\big)\in\Rset^2$ and $\bbf_{\rx}\big(\bPsi\big)\in\Rset^2$ are two-dimensional vectors describing the polarization and antenna gain of the transmit and receive antennas, respectively. For example,  $[\bbf_{\rx}\big(\bPsi\big)]_1$ and $[\bbf_{\rx}\big(\bPsi\big)]_2$ determines the received signal for the horizontal and vertical polarization, respectively, when a wave impinges on the Rx array from angle $\bPsi$. Moreover, matrix $\bM\in\Rset^{2\times 2}$ and $\bM^{(\ell)}\in\Rset^{2\times 2}$ model the possible change of the wave polarization along the LOS and $\ell$th non-LOS path, respectively, due to, e.g., different Tx and Rx array orientations or via interaction with the channel scattering objects. 

\subsection{Discussion on Channel Parameters}

The aforementioned channel models describe the structure of small-scale fading for the BS-UE, BS-RIS, and BS-UE links; however, depending on the deployment scenario, the parameters of these models (i.e., blockage, pathloss, K-factor, distances, etc.) should be generally chosen differently for each link. Some related considerations are provided~in~the~following:

\textbf{Blockage:} The BS-UE link is often assumed to be much weaker than the BS-RIS and BS-UE links, which is motivated by the following  observations. On the one hand, coverage extension to areas where the direct link is blocked is an important application use case for RIS deployment. On the other hand, for  cases where the direct link is not weak, an \textit{excessively} large passive RIS is needed for the RIS to have a non-negligible impact on the end-to-end link, which limits this application use case for the RIS deployment \cite{najafi2020intelligent}; see Fig.~\ref{Fig:PtxQ_large} in Section~\ref{sec:sim} for numerical verification of this claim. 

\textbf{LOS vs. non-LOS:} As stated before, in order to cope with the double-pathloss effect of the RIS-generated link, RISs can be deployed to have LOS connections to the BS and UEs. Since BS and RIS are fixed nodes and often deployed at greater heights, assuming a strong LOS  link with large K-factor is reasonable for the BS-RIS link. However, since UEs are often mobile nodes and are located at lower heights, the non-LOS links can be strong too which can be modeled by a small K-factor. In the case of blockage, the BS-UE links have no LOS component, i.e., $K=0$.

\textbf{Pathloss:} The distance-dependent pathloss is often parameterized as $\bar{h}_{\PL} = \beta ({d_0}/{d})^\eta,$
where $d$ is the Tx-Rx distance, $d_0$ is a reference distance at which the pathloss is $\beta$, and $\eta$ is the path-loss exponent. Since pathloss appears twice in the RIS-generated link (in both the BS-RIS and the RIS-UE channels) but once in the BS-UE link, the \textit{absolute} value of $\beta$ is quite important and normalization should be done with care. For free-space propagation in an unbounded medium, we have $\beta=(\lambda/4\pi)^2$ at $d_0=1$~m.   For instance, $\beta=-30$~dB at $d_0=1$~m assumed in \cite{wu2019intelligent} is typical for carrier frequencies below $1$~GHz \cite{aziz2014modelling}. However, for $5$~GHz carrier frequency, this value increases to $\beta\approx-45$~dB, i.e., the BS-UE pathloss  increases by $15$~dB whereas the combined pathloss of the BS-RIS-UE link grows by $30$ dB, which has to be compensated by a larger RIS.  

\textbf{Distances:} It has been shown in several previous works that it is often beneficial to deploy the RIS either close to the Tx or close to the Rx \cite{najafi2020intelligent,wu2019intelligent}. For instance, for the application use case of extending local coverage, it is reasonable to assume that the RIS is placed in close proximity of the blockage area while also maintaining a strong LOS connection to the BS. 

\section{Performance Comparison}\label{sec:sim}

\subsection{Performance Metric}

We employ the two algorithms developed in \cite{wu2019intelligent} for active  beamforming at the BS and passive beamforming at the RIS, namely the alternating algorithm in \cite[Alg.~1]{wu2019intelligent} which is initialized using the two-stage algorithm proposed in \cite[Sec.~IV-B]{wu2019intelligent}. The objective is  to minimize the transmit power, denoted by $P_\tx$, while satisfying a minimum signal-to-interference-plus-noise ratio (SINR) requirement, denoted by $\gamma_{\thr}$, at each single-antenna UE. The computational complexity of these algorithms is cubic in $Q$ due to the use of SDP, which is prohibitive for large $Q$. Therefore, we propose a scalable algorithm that borrows the concepts of tile and codebook design in \cite{najafi2020intelligent}. Thereby, the RIS is partitioned into $N_{\tile}$ tiles, where the phase-shifts of each tile, denoted by $\bomega^{(t)},\,\,t=1,\dots,N_\tile$, is taken from a pre-defined phase-shift configuration codebook, denoted by $\sM$. Let $\bh_{r,k}^{(t)\Herm}\bGamma^{(t,m)}\bH_{t}^{(t)}$ denoted the effective channel established via tile $t$ using the $m$th element of the phase-shift configuration codebook, where $\bH_{t}^{(t)}$, $\bh_{r,k}^{(t)}$,   and $\bGamma^{(t,m)}$ are respectively the BS-tile $t$ channel, the tile $t$-UE~$k$ channel, and the  reflection coefficient matrix of tile $t$ for $m$th element of the codebook. We propose to configure the RIS by iteratively configuring the tiles, where in each iteration, the phase-shift configuration  that maximizes the minimum singular value of the effective channel from the BS to UEs is chosen.  The proposed algorithm is summarized in Alg.~\ref{alg:SVD}. The complexity of Alg.~\ref{alg:SVD} is not explicitly a function of $Q$ but scales linearly with $|\sM|$. In the following, we consider a Fourier transform (DFT)-based reflection codebook and a three-bit phase wavefront codebook (cf. \cite[Sec.~III-A]{najafi2020intelligent}), i.e.,  $|\sM|=8Q/N_\tile$. This leads to an overall complexity for Alg.~\ref{alg:SVD}  that  is linear in $Q$.

\begin{algorithm}[t]
	\small
	\caption{\small Iterative Tile- and Codebook-based Algorithm}
	\begin{spacing}{1}
		\begin{algorithmic}[1]\label{alg:SVD}
			\STATE \textbf{input:} Channel coefficients $\bh_{d,k},\bH_{t}^{(t)},\bh_{r,k}^{(t)},\,\,\forall t,k$, and SINR requirement $\gamma_\thr$.
			\STATE Initialize $\bh_k=\bh_{d,k},\,\,\forall k$.
			\FOR{$t=1,\dots,N_\tile$}				
			\STATE $m^*=\underset{m\in\sM}{\mathrm{argmax}}\,\, \min\,\,\sgval(\bH^{(m)})$ where $\sgval(\cdot)$ returns the singular values of matrix $\bH^{(m)}=[\bh_1^{(m)},\dots,\bh_{N_{\UE}}^{(m)}]$ where $\bh_k^{(m)\Herm}=\bh_{k}^\Herm+\bh_{r,k}^{(t)\Herm}\bGamma^{(t,m)}\bH_{t}^{(t)}$.
			\STATE Update $\bh_{k}=\bh_k^{(m^*)},\forall k$.
			\ENDFOR		
			\STATE Given $\bh_k,\,\,\forall k$, and $\gamma_\thr$, compute the optimal BS precoder from the convex problem in \cite[Eq. (42)]{najafi2020intelligent}.
			\STATE \textbf{output:}  RIS phase shifts for all tiles and the BS precoder.
		\end{algorithmic}
	\end{spacing}
\end{algorithm}

\subsection{Simulation Setup}
We adopt the simulation setup for coverage extension depicted in \cite[Fig.~1]{jamali2022low} with the following default values for the parameters \cite{wu2019intelligent,aziz2014modelling,jamali2022low}. The BS has a $4\times4=16$ UPA whose center is  at $[30,0,10]$~m. The RIS is a UPA whose center is at $[0,50,5]$~m which consists of $N_\y\times N_\z$ square tiles along the $\y$- and $\z$-axes, respectively. The element spacing of the UPAs at both BS and RIS is half wavelength. The UEs have a single antenna and their positions are randomly generated on a $8$~m$\times8$~m square area with center $[10, 50, 1]$~m. The noise variance is computed as $\sigma_n^2=WN_0N_{\rm f}$ with $N_0=-174$~dBm/Hz, $W=20$~MHz, and $N_{\rm f}=6$~dB. We assume $5$~GHz carrier frequency, $\beta=-46$~dB at $d_0=1$~m, and $\gamma_\thr=10$. Moreover, we adopt $\eta = (3.5,2,2.8)$ and  $K=(0,10,1)$ for the BS-UE, BS-RIS, and RIS-UE channels, respectively. Except for the i.i.d Rayleigh fading model,  LOS is assumed for the BS-RIS and RIS-UE links.  For each channel, we generate $L=5$ clusters and $R=20$ subpaths (i.e., $100$ channel paths for each link or $100\times(2N_\UE+1)$ paths in total). The clusters for the BS-UE, BS-RIS, and RIS-UE channels are randomly generated in volumes $\sV_s=\{(\x,\y,\z)|\x\in[0,40]~\text{m}, \y\in[0,60]~\text{m},\z\in[0,10]~\text{m}\}$,
$\{(\x,\y,\z)|\x\in[0,40]~\text{m}, \y\in[0,50]~\text{m},\z\in[0,10]~\text{m}\}$,~and
$\{(\x,\y,\z)|\x\in[0,40]~\text{m}, \y\in[40,60]~\text{m},\z\in[0,10]~\text{m}\}$, respectively. The paths within each cluster are generated in a cube of volume $2^3$~m$^3$. The simulation results are averaged over $1000$ independent realizations of the channels and~UE~positions.

\begin{figure*}[t]
	\begin{minipage}{0.5\textwidth} 
		\centering
		\includegraphics[width=0.98\columnwidth]{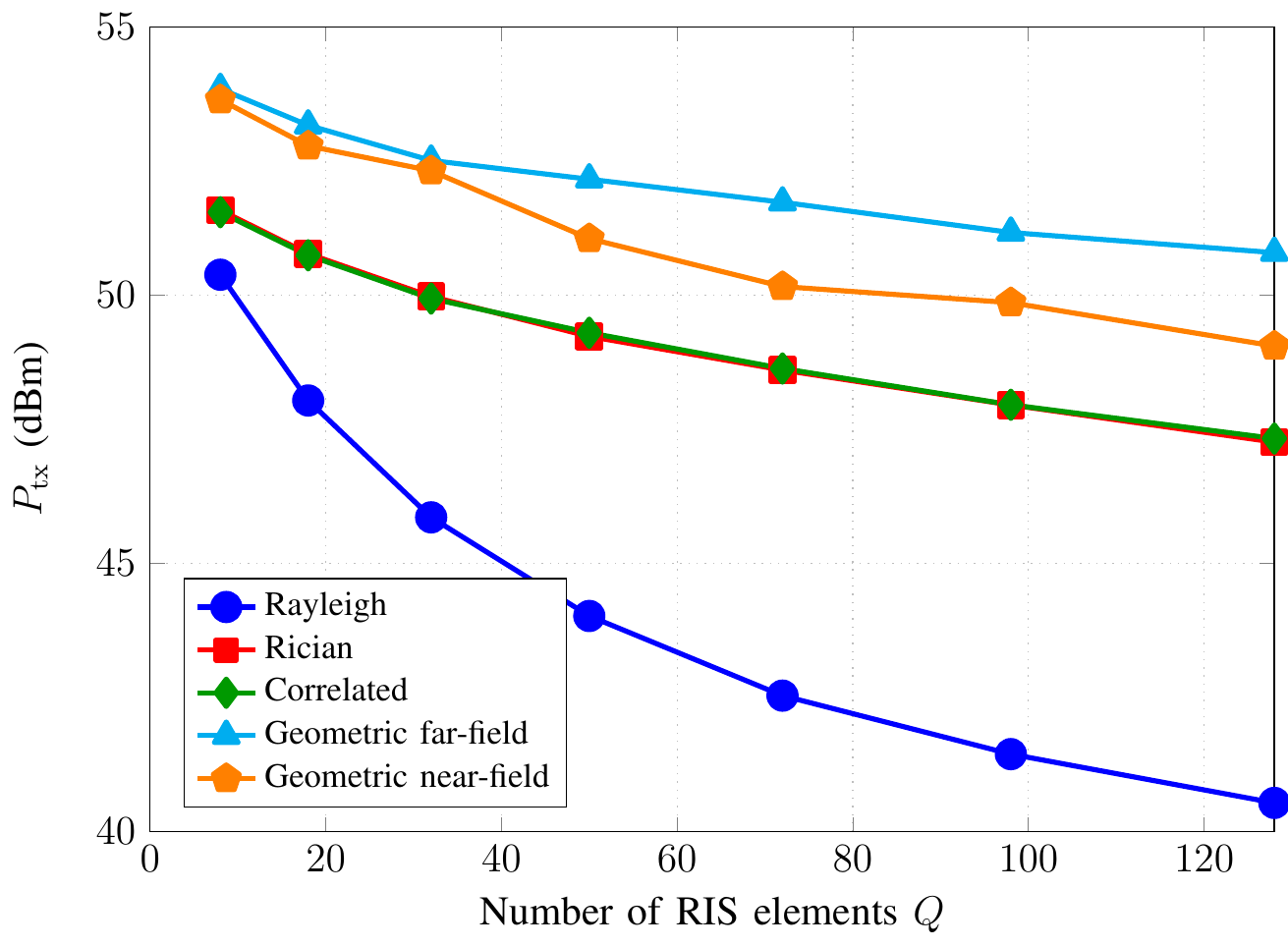} 
		\caption{$P_\tx$ vs. $Q$ for $N_{\UE}=2$, BS-UE blockage $\bar{h}_\BL=-40$~dB and algorithm in \cite{wu2019intelligent}.} \label{Fig:PtxQ_QQ}
	\end{minipage}
	\hspace{1mm}
	\begin{minipage}{0.5\textwidth}
		\centering
		\includegraphics[width=0.98\columnwidth]{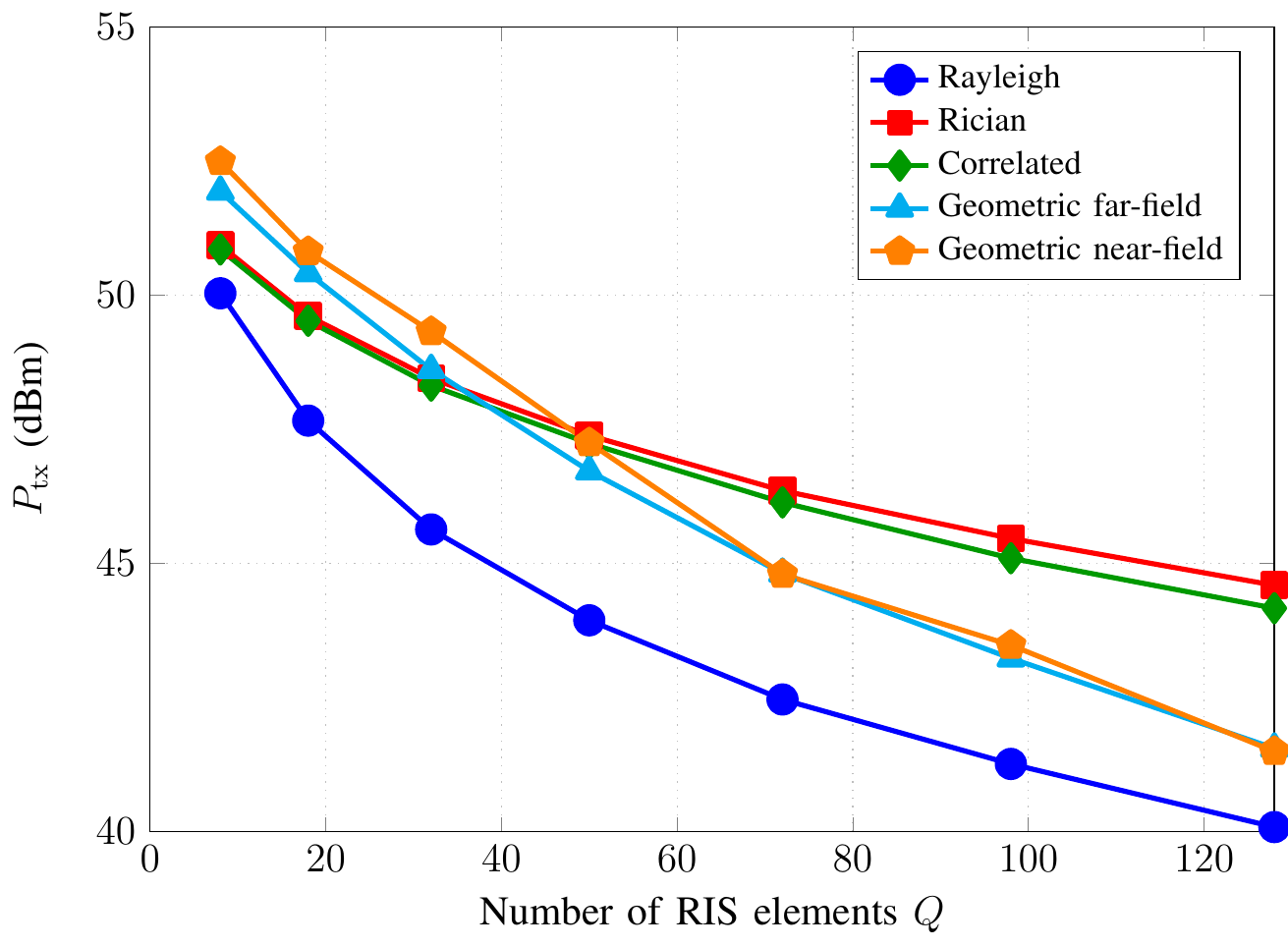} 
		\caption{$P_\tx$ vs. $Q$ for $N_{\UE}=2$, $\bar{h}_\BL=-40$~dB, $(N_\y,N_\z)=(2,1)$, and the proposed Algorithm~\ref{alg:SVD}.} \label{Fig:PtxQ_prop}
	\end{minipage}
\end{figure*} 

\begin{figure*}[t]
		\centering
		\includegraphics[width=0.7\textwidth]{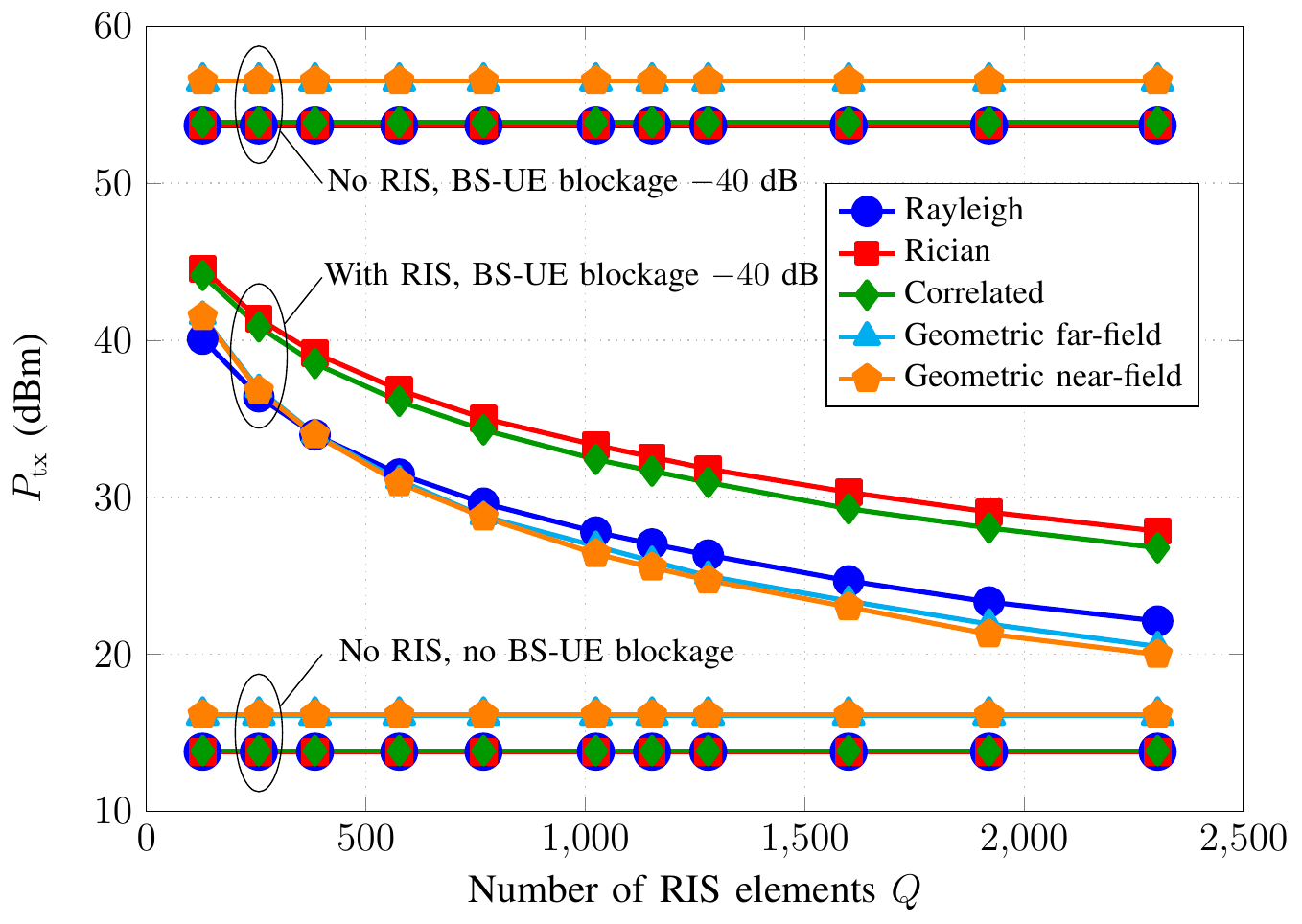} 
		\caption{$P_\tx$ vs. $Q$ for $N_{\UE}=2$, $\bar{h}_\BL=0,-40$~dB, $64$-element tile, and the proposed Algorithm~\ref{alg:SVD}.} \label{Fig:PtxQ_large}
\end{figure*} 

\begin{figure*}[t]
	\centering
	\includegraphics[width=0.7\textwidth]{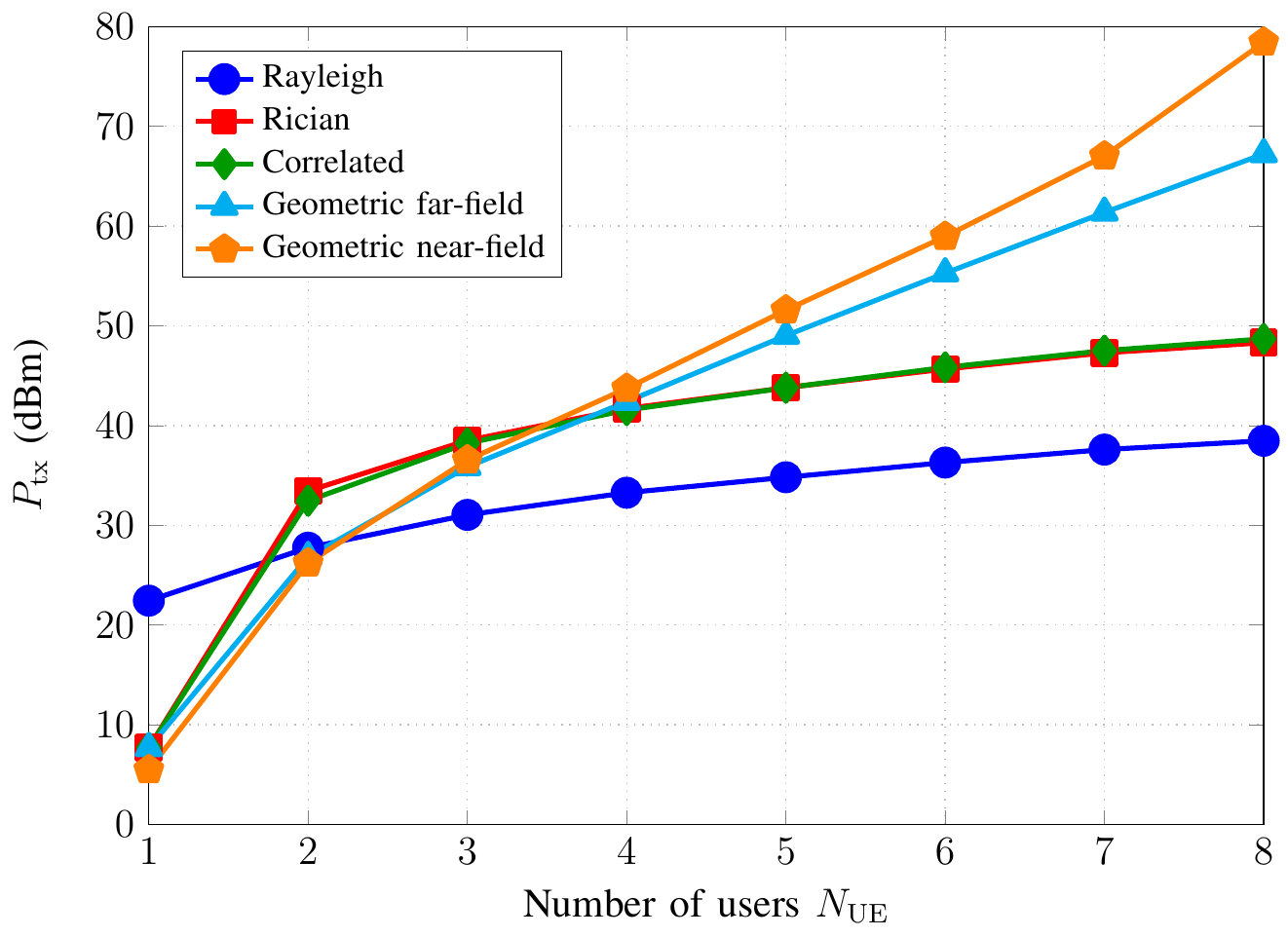} 
	\caption{$P_\tx$ vs. $N_\UE$ for $Q=1024$ ($N_\y=N_\z=4$ and $64$-element tiles), full blockage of BS-UE channel, and the proposed Algorithm~\ref{alg:SVD}.} \label{Fig:PtxK} \vspace{-0.3cm}
\end{figure*} 

\subsection{Simulation Results}

Figs.~\ref{Fig:PtxQ_QQ} and \ref{Fig:PtxQ_prop} show the required BS transmit power $P_\tx$ vs. the number of RIS elements $Q$ for $N_{\UE}=2$ when the algorithms in \cite{wu2019intelligent} and Algorithm~\ref{alg:SVD} are adopted, respectively.  Due to the high complexity of  the algorithm in \cite{wu2019intelligent}, only up to $Q=128$ RIS elements are considered.  Moreover, although a large  pathloss exponent is assumed for the BS-UE link, extensive simulation experiments have revealed that   the impact of the considered RIS is negligible unless the BS-UE link is further attenuated due to, e.g., a blockage. Therefore, we consider a weak  BS-UE link with $\bar{h}_{\BL}=-40$ dB. For this setup, Fig.~\ref{Fig:PtxQ_QQ} shows that  the required BS transmit power is significantly lower for  i.i.d. Rayleigh fading than the other models. For half-wavelength element spacing, the performance of i.i.d. Rician and correlated fading is similar which is due to the small correlation of the channel coefficients in this case.  Interestingly, while Fig.~\ref{Fig:PtxQ_QQ} shows that the algorithm in \cite{wu2019intelligent} yields the largest transmit power for the geometric fading models\footnote{The amplitude taper in \eqref{Eq:proposed} can be potentially incorporated into some of the other channel models, too. Therefore, for a fair comparison, we only include the impact of wavefront curvature in \eqref{Eq:proposed} for the reported simulations.}, Fig.~\ref{Fig:PtxQ_prop} reveals that  the proposed Algorithm~\ref{alg:SVD} requires  a much smaller  transmit power for the geometric fading models despite its lower computational complexity compared to the algorithm in \cite{wu2019intelligent}. This is due to the fact that the  Algorithm~\ref{alg:SVD} explicitly exploits the ray propagation structure of the underlying channel, i.e., the  reflection codebook  enables  reflection in the desired directions by each tile and the phase wavefront codebook controls the constructive or destructive superposition of the waves reflected by all tiles.

The results obtained with the proposed Algorithm~\ref{alg:SVD} for large RIS (i.e., $Q\geq 128$) are reported in Fig.~\ref{Fig:PtxQ_large}. In addition, for comparison, we show the results for two baselines without the RIS, namely obstructed and unobstructed BS-UE links. This figure suggests that even for the considered relatively rich scattering channel with overall $500$ paths,  several thousands of RIS elements are required to generate a link that is as strong as the unobstructed BS-UE link. 

Fig.~\ref{Fig:PtxK} shows the BS transmit power $P_\tx$ vs. the number of UEs $N_\UE$ for $Q=1024$ and  Algorithm~\ref{alg:SVD}. This figure reveals that compared to i.i.d. Rayleigh   fading, a much smaller transmit power is required for the other considered channel models when a single user is assumed. This is mainly due to the strong LOS component assumed for the BS-RIS channel which is missing in the i.i.d. Rayleigh model. However, the required transmit power significantly increases for multiple users, since in this case, the weak non-LOS components of the BS-RIS channel has to be used to convey information. Furthermore, the models based on isotropic scattering predict a moderate increase of $P_\tx$ as $N_\UE$ increases whereas the geometric channel models suggest a significant increase in the required $P_\tx$ due to the limited number of \textit{resolvable} channel paths.

\section{Conclusions}

In this paper, we have reviewed five channel models that were chosen to progressively improve the modeling accuracy for large RISs. Our simulation experiments for these channel models  have revealed important insights for system design. For instance, while for most scenarios, an idealistic rich scattering environment yields a higher system performance, the proposed RIS configuration algorithm is able to achieve a higher performance  in the finite scattering case for  large RISs  by explicitly exploiting the underlying geometric characteristics of wave propagation. Moreover, our results have shown that in realistic channel conditions, an extremely large RIS is required to approach the performance of an unobstructed direct link. Furthermore, in a channel with finite scattering objects, the number of UEs that can be simultaneously served by one RIS is quite limited.

\appendices
\section{Proof of Lemma~\ref{Lem:Correlation}}\label{App:LemCorrelation}
Under Assumptions~A1 and A2,  the distribution of each entry of $\bH^\nLOS$ approaches a Gaussian distribution as $L\to\infty$ following the central limit theorem \cite{gupta2018matrix}. A matrix Gaussian distribution, $\sMCN(\bM,\alpha\bU,\beta\bV)$, is defined 
three parameters: $\bM\triangleq\Ex\big\{\bH^\nLOS\big\}$, $\bU\triangleq\Ex\big\{(\bH^\nLOS-\bM)(\bH^\nLOS-\bM)^\Herm\big\}$, and $\bV\triangleq\Ex\big\{(\bH^\nLOS-\bM)^\Herm(\bH^\nLOS-\bM)\big\}$, where arbitrary non-negative scalars $\alpha$ and $\beta$ are chosen such that $\alpha\beta=1/\Ex\{\|\bH^\nLOS-\bM\|_F^2\}$ \cite{gupta2018matrix}. First, we have mean matrix $\bM=\bzero$, since $\Ex\{c^{(\ell)}\}=0,\forall \ell$. The second-order moment matrix $\bU$ can be computed as follows
\begin{IEEEeqnarray}{ll}\label{Eq:U}
	\bU&\triangleq\Ex\big\{\bH^\nLOS(\bH^\nLOS)^\Herm\big\} \nonumber\\
	& \overset{(a)}{=}\frac{1}{L} \sum_{\ell=1}^L  \Ex\big\{c^{(\ell)}(c^{(\ell)})^*\big\} \nonumber\\
	&\qquad\quad\times\Ex\left\{\ba_{\rx}(\bPsi_\rx^{(\ell)})\ba_{\tx}^\Herm(\bPsi_\tx^{(\ell)})\ba_{\tx}(\bPsi_\tx^{(\ell)})\ba_{\rx}^\Herm(\bPsi_\rx^{(\ell)})\right\} \nonumber\\
	& = \frac{\sigma_c^2}{L} \sum_{\ell=1}^L 
	\Ex\left\{\ba_{\tx}^\Herm(\bPsi_\tx^{(\ell)})\ba_{\tx}(\bPsi_\tx^{(\ell)})\right\}
	\Ex\left\{\ba_{\rx}(\bPsi_\rx^{(\ell)})\ba_{\rx}^\Herm(\bPsi_\rx^{(\ell)})\right\}
	\nonumber\\
	& = \sigma_c^2  \tr\{\bR_\tx\} \bR_\rx=\sigma_c^2  N_\tx \bR_\rx,
\end{IEEEeqnarray}
where equality $(a)$ follows from the statistical independence of $c^{(\ell)}$, $\bPsi_\tx^{(\ell)}$, and $\bPsi_\rx^{(\ell)}$ for $\forall \ell$.  The correlation matrix $\bR_s,\,\,s\in\{\tx,\rx\}$, is obtained in \cite[Eq.~(11)]{bjornson2020rayleigh} as
\begin{IEEEeqnarray}{ll}\label{Eq:R}
	[\bR_s]_{m,n} & = \Ex\big\{[\ba_s(\bPsi)]_n ([\ba_s(\bPsi)]_m)^*\big\} \nonumber\\
	&= \int_{\theta=-\frac{\pi}{2}}^{\frac{\pi}{2}}
	\int_{\phi=-\frac{\pi}{2}}^{\frac{\pi}{2}}
	\e^{\jj \kappa\bd^\Trans(\theta,\phi)(\bu_{s,m}-\bu_{s,n})} 
	\frac{\cos(\theta)}{2\pi} \dd\theta\dd\phi \quad\nonumber\\
	& = \sinc\big(\kappa\|\bu_{s,m}-\bu_{s,n}\|\big).
\end{IEEEeqnarray}
Similarly, the second-order moment matrix $\bV$ is obtained as $\bV= \sigma_c^2  N_\rx \bR_\tx$. Moreover, without loss of generality, we choose $\alpha=1/(N_\tx\sigma_c)$ and $\beta=1/(N_\rx\sigma_c)$ which meets the condition
\begin{IEEEeqnarray}{ll}\label{Eq:Frob}
	\alpha\beta=\frac{1}{\Ex\{\|\bH^\nLOS\|_F^2\}} 
	=  \frac{1}{\tr\{\bU\}}=\frac{1}{N_\tx N_\rx \sigma_c^2}.
\end{IEEEeqnarray}
This leads to $\alpha\bU=\sigma_c\bR_\rx$ and $\alpha\bU=\sigma_c\bR_\tx$ and completes the proof.

\bibliographystyle{IEEEtran}
\bibliography{References}

\end{document}